\documentclass{vgtc}                        %

\ifpdf%
  \pdfoutput=1\relax                   %
  \usepackage{graphicx}                %
  \DeclareGraphicsExtensions{.pdf,.png,.jpg,.jpeg} %
\else%
  \usepackage{graphicx}                %
  \DeclareGraphicsExtensions{.eps}     %
\fi%

\graphicspath{{figures/}{pictures/}{images/}{./}} %

\usepackage{microtype}                 %
\PassOptionsToPackage{warn}{textcomp}  %
\usepackage{textcomp}                  %
\usepackage{mathptmx}                  %
\usepackage{times}                     %
\usepackage{cite}                      %
\usepackage{tabu}                      %
\usepackage{booktabs}                  %

\onlineid{0}

\vgtccategory{Research}

\vgtcinsertpkg

\title{"Can You Move It?": The Design and Evaluation of Moving VR Shots in Sport Broadcast}

\author{Xiuqi Zhu\\ %
     \parbox{1.4in}{\scriptsize \centering the Future Lab, Tsinghua University \\ Northeastern University}
 \and Chenyi Wang\\ %
     \parbox{1.4in}{\scriptsize \centering the Future Lab, Tsinghua University \\ KTH Royal Institute of Technology}
 \and Zichun Guo\\ %
     \parbox{1.4in}{\scriptsize \centering the Future Lab, Tsinghua University \\ Beijing University of Chemical Technology}
 \and Yifan Zhao\\ %
     \parbox{1.4in}{\scriptsize \centering the Future Lab, Tsinghua University \\ Columbia University}
 \and Yang Jiao\thanks{corresponding author:jiaoyang7@tsinghua.edu.cn}\\ %
     \scriptsize the Future Lab, Tsinghua University}

\teaser{
  \centering
  \includegraphics[width=\linewidth]{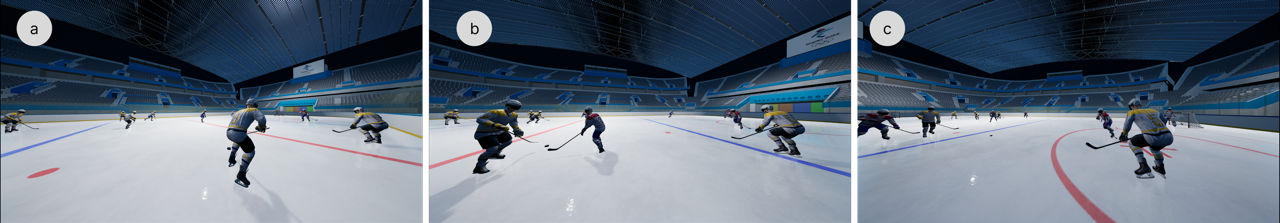}
  \caption{These three figures show three frames of the dribbling clip we designed in the VIH system, captured by a moving shot. Figure1-a, b and c are the contents of the participants as they watch the ‘pan' shot.}
  \label{fig:teaser}
}

\abstract{Virtual Reality (VR) broadcasting has seen widespread adoption in major sports events, attributed to its ability to generate a sense of presence, curiosity, and excitement among viewers. However, we have noticed that still shots reveal a limitation in the movement of VR cameras and hinder the VR viewing experience in current VR sports broadcasts. This paper aims to bridge this gap by engaging in a quantitative user analysis to explore the design and impact of dynamic VR shots on viewing experiences. We conducted two user studies in a digital hockey game twin environment and asked participants to evaluate their viewing experience through two questionnaires. Our findings suggested that the viewing experiences demonstrated no notable disparity between still and moving shots for single clips. However, when considering entire events, moving shots improved the viewer's immersive experience, with no notable increase in sickness compared to still shots. We further discuss the benefits of integrating moving shots into VR sports broadcasts and present a set of design considerations and potential improvements for future VR sports broadcasting.} %

\CCScatlist{
  \CCScatTwelve{Human-centered computing}{Human-Computer Interaction}{Empirical studies in HCI}
  
}

\begin{document}

\firstsection{Introduction}

\maketitle

Virtual Reality (VR) broadcasting has emerged as a novel form of media distribution in recent years, with numerous practical applications in large concerts, parties, and major sporting events. The utilization of panoramic 360-degree broadcasting during the 2021 Tokyo Olympics and the 2022 Beijing Winter Olympics has facilitated an immersive viewing experience, thereby enhancing the satisfaction of remote audiences who could not attend the events due to various constraints. The increasing popularity and maturity of VR equipment and content have contributed to the greater understanding and acceptance of VR viewing by audiences \cite{Kunz2019Sports, Capasa2022Mega}, consequently expanding the scope of VR's market opportunities \cite{Kletke2021Fans}.

While VR has the potential to revolutionize the way we experience sports events, there still needs to be solved in the production of VR sports content. The most significant issue is that traditional moving shots, such as pan shots and close-ups commonly used in broadcast, film, and television cannot be presented in VR broadcasting due to the 360-degree camera could not move, thereby hindering the immersive and dynamic viewing experience inherent to traditional audio-visual language. Although current VR photography technology cannot achieve moving or mixed camera shots in sports games, locomotion in VR games \cite{Tseng2022Headwind}, films \cite{Li2021Movie}, and simulators \cite{Matviienko2022Bicycle, Katsigiannis2019QoE}, have been widely explored. Previous research suggested that the incorporation of motion camera experience enhances the content's narrative and fosters a more three-dimensional viewing experience, including the sense of co-presence \cite{Li2021Movie} and immersion \cite{Boletsis2019Locomotion}, but also reveals that users' movement in VR environments may result in VR sickness \cite{Chang2020Review, Clifton2020Effects}. In this work, we suggest a hypothesis that incorporating motion shots into VR sports broadcasting, coupled with appropriate content design to manage users' VR sickness, could comprehensively enhance the viewing experience by compensating for the lack of aesthetic elements in the current VR experience, thus delivering a rich and immersive viewing experience for the future broadcast.

The primary aim of this paper is to investigate the viewer's experience of viewing moving shots in VR sports event broadcasting, with a specific focus on ice hockey. To begin, we conducted a field study involving observing a VR hockey broadcast test game and interviews with broadcast production experts to comprehend the current challenges associated with VR broadcasting. Our formative findings revealed a question among production staff and researcher regarding the strategy, advantage and approach behind designing moving VR shots instead of still shots.

In response to these challenges, we first proposed an event segmentation theory to divide the entire ice hockey game into distinct clips. Furthermore, we developed a digital twin space, the Virtual Ice Hockey (VIH) system, which incorporated four motion shots based on the fundamental audio-visual language of television and film and a still shot, serving to display each Cinematic VR (CVR) clip \cite{Mateer2017CVR}. Subsequently, we conducted two rounds of user experiments and collected data through capture   participants' perspectives, questionnaires, user interviews, wherein participants viewed single CVR clips and entire event CVR clips, respectively. 

Our findings indicate that while there is no significant difference between moving shots and still shots in single clips, the incorporation of moving shots significantly enhances the overall immersive experience of viewers in VR sports broadcasts compared to still shots, with no notable increase in VR sickness. We further discuss the benefit of integrating moving shots in VR sports broadcast and offers valuable insights for evaluating and designing moving shots in future research and production of VR sports broadcasts.

This paper contributes to the HCI community in two ways. Firstly, we report on the issues with current VR hockey broadcasts and explore how they can be explored using digital twins. Secondly, we designed various moving shots for VR sports broadcasts from single clips to entire events and evaluated the audience's viewing experience to suggest the movement of VR shots in sports broadcasts.

\section{Related Work}
\subsection{VR Sports Broadcast}
Sports event broadcasting plays a crucial role in expanding the influence of sports events, disseminating event information, and meeting the demands of sports consumption \cite{Turner2007Impact,boyle2002new}. In recent years, with the rapid development and commercialization of VR technology, many broadcasters have embraced this medium for sports event broadcasts \cite{Iyer2022Qatar,Kletke2021Fans,Kunz2019Sports,Uhm2020neurophysiological}. Research indicates that audiences are increasingly receptive to this novel form of viewing experience \cite{Kunz2019Sports,Capasa2022Mega}. As sports broadcasting continues to gain popularity, audience expectations for immersive viewing experiences and enhanced services are on the rise. The emergence of VR broadcasts reflects spectators' preference for experiencing the game as if they were physically present, surpassing the traditional television viewing experience \cite{Ludvigsen2010Spectator,Ko2011quality}. For example, Daehwan et al. proposed that VR's primary function is to provide an immersive experience to sport media consumers by enhancing tele-presence \cite{Kim2019Flow}.

Nevertheless, current VR sports broadcasts face challenges related to user adoption, business models, and content production \cite{Kletke2021Fans}. These issues are typical for an industry that is still in its early stages, yet the advantages of VR sports broadcasting are becoming increasingly evident to viewers and related industries. With VR technology, spectators can enjoy a realistic experience of mega sports games while saving time and reducing costs \cite{Uhm2020neurophysiological,Kim2019Flow,Plante2003Mood}. Furthermore, the advent of VR has opened up new revenue streams for professionals in the field, including coaches, athletes, and broadcasters, who have already started to benefit from VR's development.

\subsection{The Experience of Watching Moving VR Shots}
Locomotion is an important component of VR since it can strongly influence user experience \cite{Tseng2022Headwind,bowman20043d,Bozgeyikli2016Teleport}, which is also defined as self-propelled movement in virtual worlds \cite{McMahan2015Handbook}. In previous studies, two common locomotion techniques, teleportation, and continuous locomotion, were used as interaction methods in VR games \cite{Tseng2022Headwind}, movies \cite{Li2021Movie} and simulators \cite{Matviienko2022Bicycle,Katsigiannis2019QoE}. 

Virtual Reality Sickness (VRS)\cite{Chang2020Review} was also known as Visually-Induced Motion Sickness (VIMS)\cite{Hettinger1992Induced}, Cyber Sickness(CS)\cite{McCauley1992Cybersickness}, or Virtual Simulator Sickness\cite{Howarth1997VSSS}. We collectively referred to these negative effects users experience during or after immersion into virtual reality as VR sickness in this paper \cite{Chang2020Review,Kim2015Oculus,Kohl1983Neurconflict}. For VR locomotion, sensory conflict induced by the disparity in motion between two sensory systems – visual and vestibular is inevitable \cite{reason1975motion,Kohl1983Neurconflict}. Previous literature illustrated that, compared to teleportation, continuous locomotion allowed the viewer to move continuously, which usually caused significant VR sickness \cite{Clifton2020Effects,Chang2020Review,Caserman2021Cybersickness,Vlahovic2018QoMEX}. By contrast, teleportation locomotion teleports the user from the current location to the destination \cite{Tseng2022Headwind}, which also brings low VR sickness \cite{Vlahovic2018QoMEX}. However, the cause of VR sickness is related to hardware, content, and human factors (i.e., age, gender, and VR related experience) \cite{Chang2020Review}. Therefore, concluding which locomotion technique is better is generally filled with limitations.

Nevertheless, some studies suggested that teleportation may significantly reduce immersion \cite{Boletsis2019Locomotion} and increase spatial disorientation for VR viewers \cite{Bowman1997Travel,Bakker2003Head-Slaved}. Immersion is a very important and special feeling that VR brings to the experience. For VR, a person immersed in a virtual environment (VE) may identify with his or her virtual body (VB) and experience a sense of presence if their senses confirm that the VB is functioning effectively within the VE \cite{Tromp1995Presence}. When the user feels a stronger sense of presence, a stronger sense of immersion is created, which define as in VE, the user interacts with the VE in some way and temporarily feels that this state is real \cite{Slater1994Depth}. In addition, for spatial disorientation, VR producers focus on how to attract the viewer's attention through various methods. Cagri et al. first designed and implemented a test environment for VR attention models inspired by various Visual Attention Models (VAMs) applied in film and television to collect the viewport trajectory when participants watched omnidirectional video \cite{Ozcinar2018Attention}. Their results indicated that viewers do not pay different visual attention to the same content repeatedly, and this amount depends on the complexity of the camera movement of the omnidirectional video. Overall, viewing the content in VR in a locomotion way is a holistic experience. Although continuous locomotion in VR may mostly give viewers a stronger sense of VR sickness, continuous movement of refined quality will also bring users a more intense sense of immersion and accurate spatial awareness.

In this paper, we mainly explore the experience of watching different moving shots in VR environment. Thus, based on our experiences and literature reviews of moving shots in sports broadcasts, we choose continuous locomotion techniques, a process-oriented technique, as design strategies for VR moving shots \cite{Bozgeyikli2016Teleport}. 
This kind of moving shots provide opportunities for sectors easier focus on the content of process in VR sports broadcasts. Thereby, we further explore the immersion, VR sickness and content expression of different moving shots.

\subsection{Audio-Visual languages in VR Sport Broadcast}
Audio-visual language is the means of expression of all video works. The VR video audio-visual languages was developed by the application of elements such as images, shots, and film editing. Image is the basic vocabulary of VR artistic language, film editing connects VR video like grammar and the shots offer the context \cite{Peng2017Artistic}. The shots of virtual videos can be categorized as still shots, moving shots, and autonomous shots. In the current VR sports broadcasts, the most commonly used footage is the still camera shots and there are relatively few applications of moving shots. For example, in the 2018 Pyeongchang Winter Olympics, there were multiple panoramic cameras installed on the cross-country skiing course. Although the practice of VR sports broadcasts is available, the theoretical content is not sufficient. However, VR sports broadcasting still follows the content of the audio-visual language of VR video. Therefore, by analyzing the characteristics of VR video, we can provide a reference for VR sports broadcasting. 

VR videos are emerging as a medium for both self-exploration and the establishment of social identity via bullet comments \cite{Vosmeer2017Orpheus,Li2022Bullet}. The experience of watching VR video is not entirely passive nor fully active \cite{Vosmeer2017Orpheus}. Thus, many current research focus on how to design a good VR video or cinematic VR (CVR). This medium lies in between traditional cinema and VR \cite{Pillai2019Storytelling}, thus the audio-visual language of VR video could be based on traditional cinematic language such as cognitive event segmentation and it could also offer new iterations in expansive visual technologies \cite{Leotta2018Touring}. We summarized three key factors could affect the viewing experience when designing a CVR based on piror studies. (1) The height: people generally prefer lower camera heights to higher camera heights, and the VR video guidelines are recommended to place the camera at the head height \cite{Rothe2018Height,passmore2017literacy360}. (2) The distance from the object to the camera: People would have a more intense feeling when the object in the video stays around them than in the distances \cite{Probst2021Distances}. However, it would cause a negative effect on the experience when people are very close to the camera \cite{Keskinen2019Effect}, so the ideal camera placement distance is between 2m and 3m \cite{Brown2016Attention}. (3) The editing: Editing techniques include the techniques based on film/television and VR features \cite{Husung2019Orbs}. The most common technique is ‘Fade’ because it meets the audience's psychological expectations \cite{Dooley2019Proximity}. Regarding the timing of editing, the Probabilistic Experiential Editing (PPE) proposed by Jessica Brillhart is currently the most accepted articulation point approach \cite{Jessica2016Blink}. Jiang et al. recently introduced a new deep-learning framework for camera keyframing, enabling customized and automated video generation in virtual environments. They also provided a camera trajectory editing interface to support editors in managing timelines, characters, keyframes, and previews\cite{Jiang2021Keyframing}.

These characteristics of VR video would affect the viewer's experience to a certain extent. Based on these reviews and our knowledge, we believe these could provide experience and reference for VR Sports Broadcast. Thus, we followed these suggestions and guidance to design the moving shots in VR sports broadcasts based on the traditional audio-visual languages in our experiments.

\section{System Design}
\subsection{Field Study}
Sports event broadcasting is a crucial way to expand the influence of sports events, disseminate event information and meet the derived sports consumption. Prior to our investigation, we conducted a comprehensive field study to explore several key objectives: (1) identify current challenges in VR ice hockey broadcasting, (2) assess the design and extent of how VR ice hockey broadcasts deliver a positive experience for the audience, and (3) explore the potential for incorporating moving shots in VR ice hockey broadcasts.

To address these questions, we conducted observations at an ice hockey stadium, where both traditional cameras and VR cameras were strategically positioned around the stands. Additionally, we engaged in discussions with experts in communication signals, VR production, and ice hockey coaching. Through these interactions, we found that regarding the current state of VR broadcast shots in ice hockey are still, as VR cameras lack mobility. This limitation hampers the dynamic nature of the broadcasts. The non-portability and high costs associated with the production process of VR sports broadcasts also pose significant challenges. Moreover, the fast-paced nature of ice hockey games presents difficulties in designing and evaluating comprehensive audio-visual language broadcasts, even if capturing moving VR shots becomes feasible.

\subsection{Event Segmentation Theory}
Based on our field study, we believe that simplifying the content of VR broadcasts for ice hockey can enhance our study and analysis. To achieve this, we introduce the event segmentation theory. Inspired by Vladimir Propp's work, we suggest that narrative events can be meticulously structured around the concept of ‘clip’ and ‘round’ of athletes' actions \cite{propp1968morphology}. The clip represents the fundamental narrative unit, while the round refers to a entire unit like a hockey game. By identifying key clips such as collision, defense, passing, hitting, dribbling, and tactical formation within a round, we can focus on important moments amidst the numerous actions in a hockey game.

\subsection{Implementation}
\begin{figure}[h]
    \includegraphics[width=\linewidth]{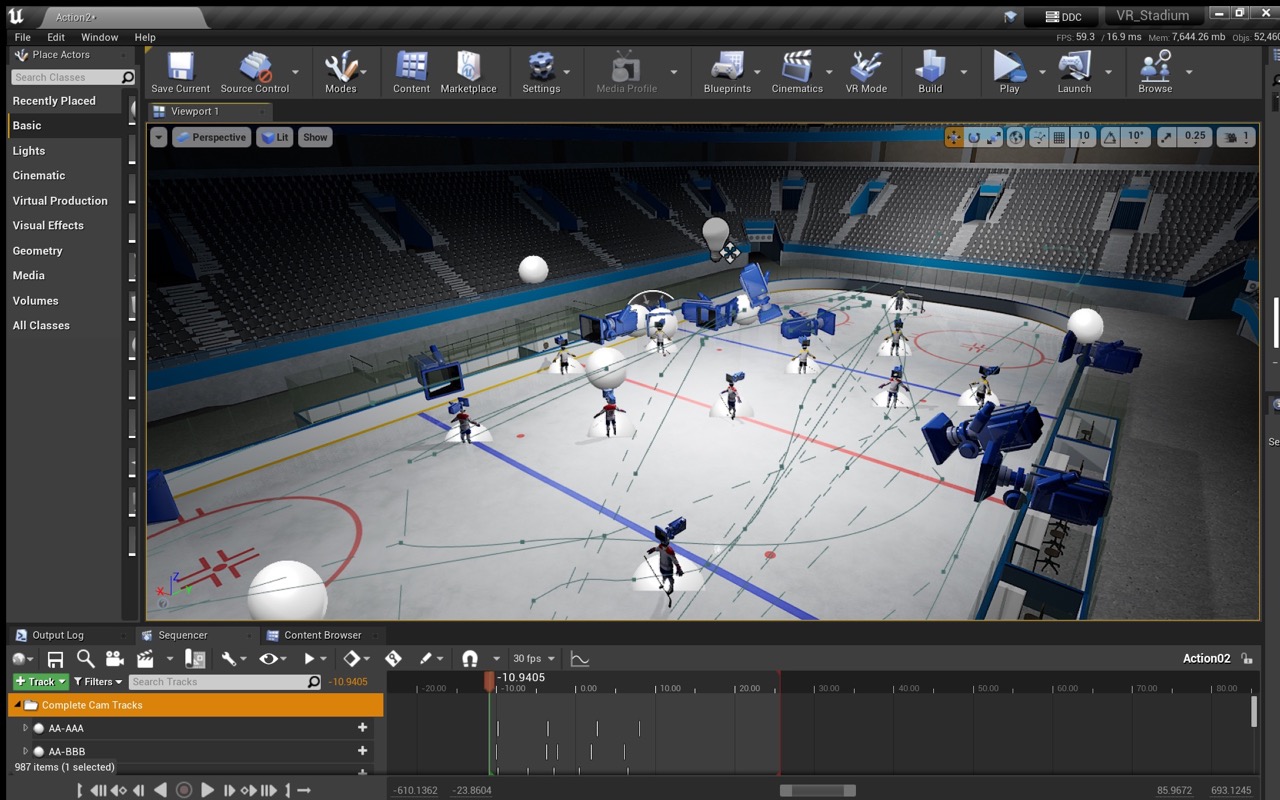}
    \caption{The interface and overview of VIH system.}
    \label{interface}
\end{figure}

To address the identified issues and difficulties from our field study, we believe there is much theoretical and exploratory work to be done before deploying VR moving shots in the field. Thus, we developed a Virtual Ice-Hockey (VIH) system within a digital twin environment using \textit{Unreal Engine 4} (see Figure \ref{interface}). The VIH system encompasses three key clips: (1) enabling virtual athletes to perform predetermined movements, (2) allowing multiple 360-degree cameras to traverse along custom tracks at variable speeds, and (3) facilitating the selection of specific CVR clips for viewing.

To implement these clips, we utilized blueprints to configure multiple virtual cameras and govern the playback of CVR clips. Additionally, we collaborated with ice hockey players to capture video data of their actions, employing video motion capture techniques to extract skeletal movements. These movements were then applied to athlete models in \textit{Blender} to generate a comprehensive set of ice hockey actions. Importing the action set into \textit{Unreal Engine 4}, we entrusted hockey experts with designing the virtual ice hockey event.

\section{Study 1}
Our primary research question aims to comprehend the impact of moving shots in VR ice hockey broadcasting. Therefore, in the initial study, our focus was to investigate whether moving shots could enhance the viewing experience for the audience, while also determining the optimal approach to designing these shots within a single clip. Subsequently, we delve into the performance of moving shots throughout the entire event in the second experiment.

\subsection{Participants}
We enlisted 12 participants (10 female, 2 male) recruited through social media and word of mouth, with ages ranging from 21 to 29 (M=24.16, SD=2.64). Participants were requested to disclose their frequency of VR technology usage and any experiences of 3D vertigo or motion sickness before the experiments. Among the participants, ten had no prior experience with VR, while only two had limited exposure to VR on a few occasions. None of the participants reported experiencing 3D vertigo in their self-reports. All participants possessed normal or corrected-to-normal vision. Monetary compensation was provided to participants, equivalent to the time dedicated to the experiment.

\subsection{Study Design}
\begin{figure*}[h]
    \centering
    \includegraphics[width=\linewidth]{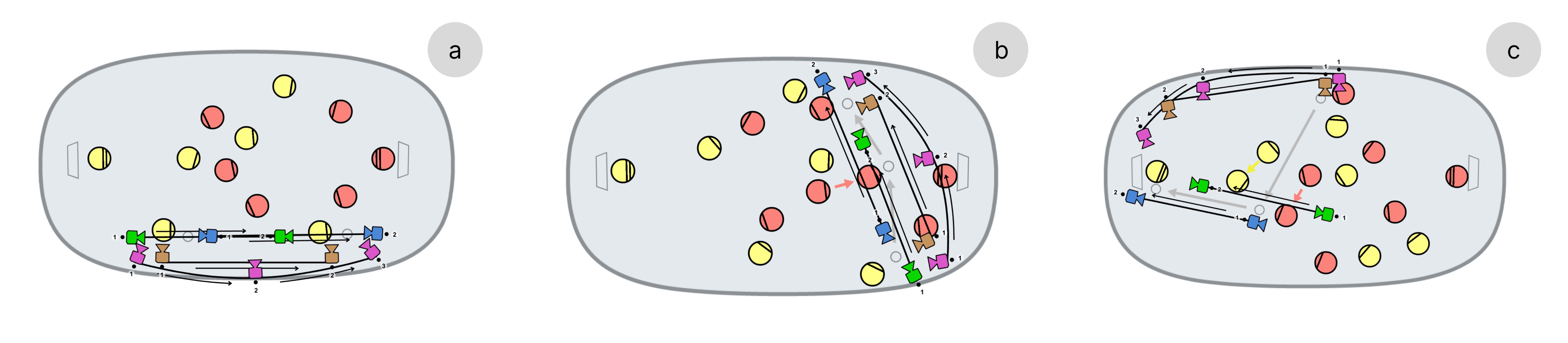}
    \caption{These three figures demonstrate the four shots movement tracks based on 2D audio-visual language and the athletes' movement tracks in three clips, including (a) dribbling, (b)stealing, and (c)shooting. The yellow and red circles represent the athletes of the two teams, with six on each side. The four colors of the camera tracks represent different audio-visual languages, green for ‘Track-in’, blue for ‘Track-out’, pink for ‘Pan’, and brown for ‘Dolly’. }
    \label{fig3}
\end{figure*}
In the present study, we constructed an event following the principles of event segmentation theory. This event dubbed a ‘defense-offense transition’, was simulated as a round in the VIH system. This simulation incorporated three distinct clips: dribbling, stealing, and shooting. For each of these clips, we designed five diverse CVR shots. These consisted of four moving shots and one still shot to examine the viewers' engagement and experience. The design strategies behind these shots are elucidated in Figure \ref{fig3}.

\textbf{1) Capturing CVR Aesthetics:} The aesthetic design of our CVR shots was inspired by traditional ice hockey television broadcasts to represent the event while ensuring visual appeal accurately. \textbf{2) Mitigating VR Sickness:} VR sickness, often a result of VR movement and other hardware-related factors \cite{Chang2020Review}, is an unavoidable concern. To address content-related causes of this issue, we established the camera zone at a personal distance and incorporated full body shots \cite{Keskinen2019Effect}. Furthermore, we introduced slow start/stop indexing and fade-in/fade-out effects for each moving shot. \textbf{3) Adhering to Audio-Visual Language Principles in Film and Television:} In CVR, viewpoint/point-of-view surpasses the limitations of traditional shots, offering viewers the freedom to explore the scene \cite{Cherni2020Review}. Consequently, our shot design incorporates four foundational audio-visual movements (track-in, track-out, pan, and dolly) following the guiding principles of film and television.

To augment the realism of our CVR shots, we integrated three audio clips, each featuring different content, such as crowd cheers, player movements, and puck strikes. We crafted three distinct moving shots for each clip, their trajectories aligned with four different methods of virtual camera motion. Subsequently, film and hockey experts were invited to review and select the shots that best encapsulated the ice hockey viewing experience. Overall, we curated 15 shots across the three clips, each lasting approximately 7 seconds. 

The \textit{‘Track-in shot’} exhibits a linear trajectory with no camera rotation, moving forward as the event unfolds. 2) The \textit{‘Track-out shot’} shares a similar motion track with the ‘Track-in shot’, except it moves backward, constantly facing the athlete. 3) The \textit{Pan shot’} follows a curved path, moving laterally relative to the athlete. 4) The \textit{‘Dolly shot’} advances along a parallel path beside the athlete. 5) For the solitary \textit{‘Still shot’}, we positioned it in the front row of the stadium stand, mimicking the perspective offered by prevalent VR live broadcast shots.

\begin{figure}[h]
    \centering
    \includegraphics[width=\columnwidth]{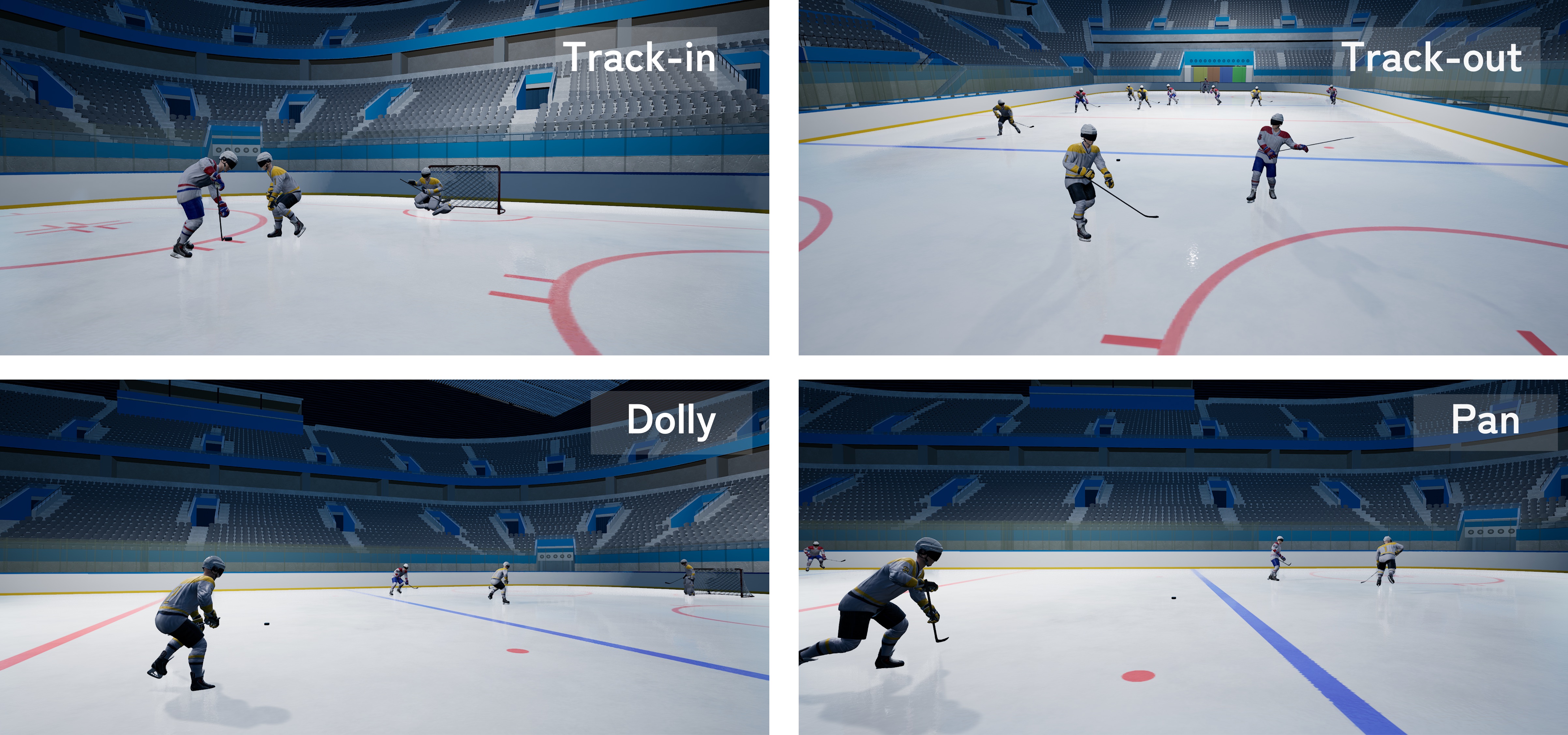}
    \caption{These four pictures depict four types of traditional audiovisual language based on the movement of the VR camera. From top to bottom and left to right, they are ‘Track-in’, ‘Track-out’, ‘Dolly’, and ‘Pan’.}
    \label{fig:enter-label}
\end{figure}

\subsection{Materials}

The experiment was conducted within a 4x4 square meter area. Participants were equipped with an HTC Vive Pro2 Head-Mounted Display (HMD) and seated in a mobile chair. Two base stations were positioned diagonally to ensure stable signal transmission. A laptop, placed on a nearby circular table, was used for participants to complete the questionnaire (refer to Figure \ref{set-up}). The objective of this study was to undertake a comprehensive evaluation to analyze the viewing experience facilitated by different VR shots. Accordingly, we collected data on the following dimensions:

\begin{figure}[h]
\centering
\includegraphics[width=\columnwidth]{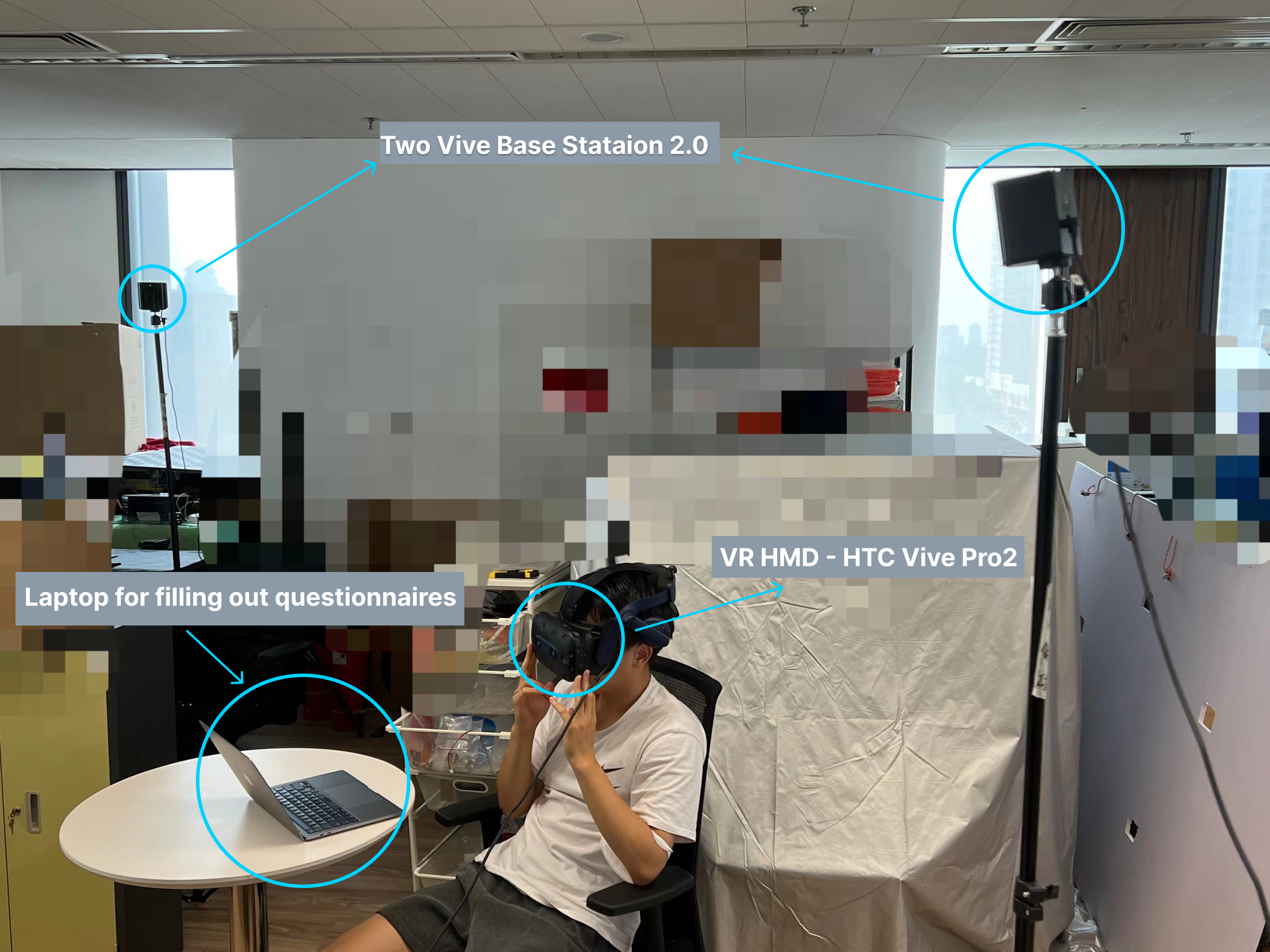}
\caption{The experimental setup in a 4x4 square meter area.}
\label{set-up}
\end{figure}

\begin{figure*}[t]
    \centering
    \includegraphics[width=\linewidth]{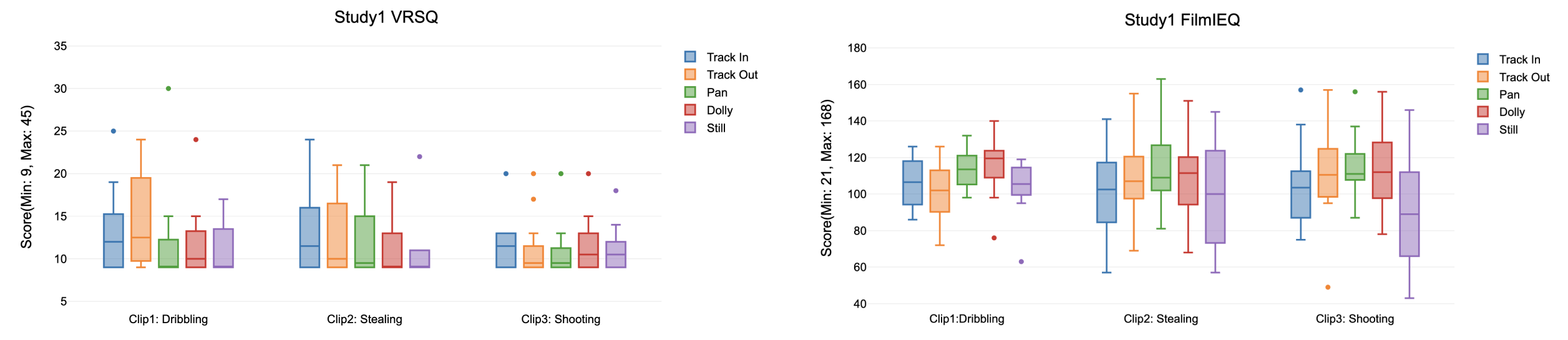}
    \caption{These two box-and-whisker plots illustrate the distribution of total VRSQ \textit{(left)} and FilmIEQ \textit{(right)} scores across five shots in three clips, including clip1: dribbling, clip2: stealing, and clip3: shooting.}
    \label{fig7}
\end{figure*}
\textit{Virtual Reality Sickness:} Participants were asked to complete the Virtual Reality Sickness Questionnaire (VRSQ)\cite{Kim2018VRSQ} after viewing each shot. The VRSQ consists of nine symptoms, including general discomfort, fatigue, eyestrain, difficulty focusing, headache, fullness of the head, blurred vision, dizziness (with eyes closed), and vertigo. To quantify the participants' discomfort, we utilized a Likert scale ranging from 1 (‘I do not experience this symptom at all’) to 5 (‘I experience this symptom intensely’). As per \cite{Kim2018VRSQ}'s instructions, VRSQ scores ranged from 9 (‘I do not experience VR sickness at all’) to 45 (‘I experience intense VR sickness’).

\textit{Immersive Experience:} After each shot, participants were requested to fill out the Immersive Experience Questionnaire for Film and TV (FilmIEQ)\cite{Rigby2019FilmIEQ} in order to evaluate the aesthetics and viewing experience of this shot. To gauge the immersive experience quantitatively, we employed a Likert scale that ranged from 1 (‘I do not experience this feeling at all’) to 7 (‘I experience this feeling intensely’). The FilmIEQ includes 24 questions across four factors: captivation, real-world dissociation, comprehension, and transportation. As per \cite{Rigby2019FilmIEQ}'s instructions, the scope of FilmIEQ scores ranged from 24 (‘I am not at all immersed in this CVR clip’) to 168 (‘I am deeply immersed in this CVR clip’).

\textit{Participants' Perspective:} For each shot, we captured the participants' perspectives at regular intervals. Every 0.5 seconds, a frame from the participant's viewpoint was recorded. These frames were subsequently analyzed to determine whether the participants' viewing aligned with our predefined instructions.

\subsection{Procedure}
The study commenced with a detailed introduction of the experimental procedure to the participants, which was succeeded by obtaining their informed consent and gathering demographic data. Subsequently, participants were outfitted with the Head-Mounted Display (HMD) and viewed five shots with randomized order without interruption. After the initial viewing, participants were asked to review the sequence of five shots one by one. After each shot, participants removed the HMD and completed a questionnaire on a provided laptop. This approach was designed to minimize cognitive biases associated with the novelty and potential discomfort of the initial viewing experience \cite{Moghadam2017Scene}.

The same viewing process, including the randomization of shots during the second viewing, was carried out for the following two clips. A 5-minute break separated each clip viewing to decrease fatigue \cite{Putze2020Breaking}. Participants were further requested to share brief impressions of the CVR clips after each viewing session.

After the viewing sessions, a brief 15-minute interview was conducted with each participant to gain deeper insights into the shots' design and their viewing experience and perceptions. Consequently, the complete experimental procedure, including the interview, lasted approximately one hundred and five minutes per participant, with participants exposed to the VR environment for at least 10 minutes.

\subsection{Result} 
In this study, 12 participants completed the VRSQ and the FilmIEQ. We captured and analyzed their perspectives throughout the experiment, and we recorded some key user voices and insights, which are reported in the subsequent discussion sections. In addition, we performed descriptive data analyses on the collected variables and illustrated the results in graphical form. We detail the findings for each variable in the subsequent paragraphs.

\textit{Virtual Reality Sickness:} Our analysis revealed that ‘Track-in’ and ‘Track-out’ shots consistently generated higher VRSQ scores than the remaining shot ‘Pan’, ‘Dolly’, and ‘Still’ across various clips. The sole exception was clip3, in which shots3-4 scored higher than shots3-1 and shots3-2, as depicted in Figure \ref{fig7}-left. Despite these variations, the one-way ANOVA test results indicated no statistically significant differences among the shots within each clip—clip1 clip1 (F(4, 55)=0.95, p\textgreater0.05), clip2 (F(4, 55)=0.63, p\textgreater0.05), and clip3 (F(4, 55)=0.13, p\textgreater0.05). The average VRSQ scores for the 15 shots ranged from 10.67 to 14.83(M=12.00, SD=4.16), suggesting that none induced significant VR sickness among the participants. This outcome supports that our shot movement designs are suitable and easily tolerated.

\textit{Immersive Experience:} Our analysis revealed that the ‘Pan’ and ‘Dolly’ shots consistently achieved higher FilmIEQ scores than the ‘Track-in’ and ‘Track-out’ shots across all clips, as depicted in Figure \ref{fig7}-right. Furthermore, all four types of moving shots exhibited superior FilmIEQ scores compared to the single still shot in each clips. The average FilmIEQ scores for the 15 shots ranged from 91.17 to 115.42, with a mean (M) of 107.01 and a standard deviation (SD) of 22.19. However, the one-way ANOVA test indicated no statistically significant variations among the different shots in clip1 (F(4,55) =2.31, p\textgreater0.05), clip2 (F(4,55) =0.66, p\textgreater0.05) and clip3 (F(4,55) =0.13, p\textgreater0.05).

\textit{Participants' Perspective:} We captured approximately 15 frames of participants' perspectives for each shot. Our analysis of these frames suggested that they could be amalgamated into a complete depiction of the clips. Each frame contained relevant information (i.e., players and pucks), with only a few frames displaying less pertinent content (i.e., ceilings and bleachers). These findings imply that our shot movement designs are easy to follow and understand. Thus, we did not conduct further quantitative analysis of these frames.

\section{Study2}
Going a step further to understand our primary research question, we following conducted a study to explore the impact of moving shots throughout the entire event. 

\subsection{Participants}
We recruited another group of 12 participants (5 female, 7 male) via social media channels and personal referrals, aged between 20 and 29 (M=23, SD=3.1). Each participant was asked to provide information on their frequency of VR technology usage, as well as any previous experiences of 3D vertigo or motion sickness, prior to the commencement of the experiments. Among the participants, two had no previous VR experience, six had sporadic exposure to VR, and four reported extensive experience with VR usage or development. Notably, none of the participants disclosed any instances of 3D vertigo in their self-reports. All participants possessed normal or corrected-to-normal vision. Commensurate with the time dedicated to the experiment, compensation was provided to all participants. 

\subsection{Study Design}

\begin{figure*}[h]
    \centering
    \includegraphics[width=0.7\linewidth, height=0.17\textheight]{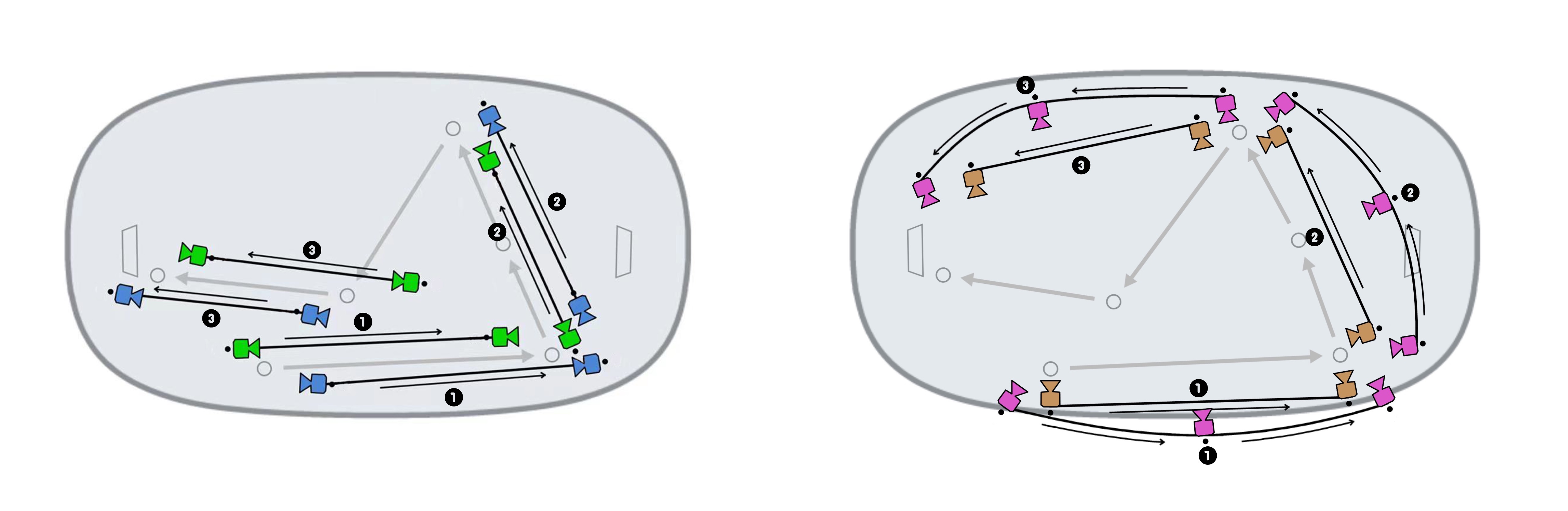}
    \caption{These two figures demonstrate four combined shots movement tracks based on 2D audio-visual language and Study1 Design, with the athletes' movement tracks for a defense-offense transition event. The three colors of the camera tracks represent different audio-visual language, green for ‘Track-in’, blue for ‘Track-out’, pink for ‘Pan’, and brown for ‘Dolly’.}
    \label{fig7}
\end{figure*}

In this study, we further designed five different shots (track-in, track-out, pan, dolly, and still) to explore and analyze their watching experience across the entire CVR event. Since we aim to extend the previous CVR camera tracks in Study1, we combined the three clips from study 1 to form a entire event clip with hard cuts in the editing. The complete camera tracks are illustrated in Figure\ref{fig7}. Specifically, the ‘still’ shots are strategically maintained in their original position from Study 1, specifically in the front row of the stadium stand. This position was chosen to mimic the perspective of prevalent VR live broadcast shots without editing throughout the clip. To minimize the VR sickness of our shots, we used hard-cut as our transition method between clips and allowed participants freely control their point of view by only giving their starting angle \cite{Dooley2019Proximity, Moghadam2017Scene}. Subsequently, we solicited feedback from experts in hockey and a cinema director to review the content and shot movement, ensuring that our combined clips offered clear storytelling and were easy to follow. They first emphatically affirmed our shot design and editing methods.
Additionally, they recommended aligning our cuts with key moments in ice hockey, such as hits and passes. They also suggested adding sound effects when the hockey puck was hit to amplify the impact of the cuts. We modified the shots according to their suggestions and used them in the following experiments. Overall, we curated 5 combined shots, each lasting approximately 30 seconds. 

\subsection{Materials and Procedure}
In the present study, we employed the same experimental apparatus and questionnaires as in Study 1. As observed in Study 1, the majority of participants accurately followed our shot directions, and their viewing angles predominantly conformed to our design. Consequently, additional perspective captures were deemed unnecessary in this study.

During this study, participants were firstly requested to provide demographic information such as their age, field of study, and any pre-existing conditions like VR-induced vertigo or motion sickness. After an introduction and explanation of the process, participants viewed the complete set of five shots initially, followed by a second round where they experienced a randomized sequence of the same shots. This sequencing rationale is aligned with the procedure outlined in Study 1. During the second round of viewing, participants were instructed to remove the HMD and subsequently complete the pertinent questionnaires using a laptop, each time after the appraisal of a given shot. The total duration of the experiment was around 40 minutes, with participants exposed to the VR environment for at least 5 minutes.

\subsection{Results} 
\begin{figure*}[h]
    \centering
    \includegraphics[width=\linewidth]{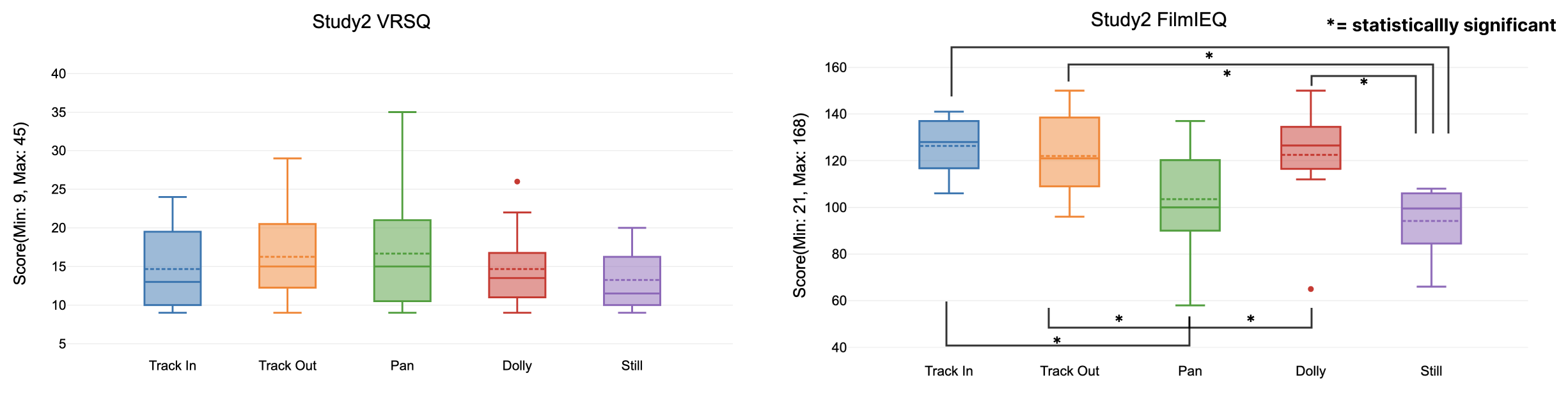}
    \caption{These two box-and-whisker plots illustrate the distribution of total VRSQ \textit{(left)} and FilmIEQ \textit{(right)} scores across five shots in the combined clip. The * represents the statistically significant difference between the two shots.}
    \label{fig8}
\end{figure*}

\textit{Virtual Reality Sickness:} In the present study, 12 participants completed both the initial VRSQ and the subsequent FilmIEQ questionnaires. The VRSQ results revealed marginal variance in the mean scores of the five shots. Specifically, the ‘Track-in’ shot (M=14.67, SD=5.44) and the ‘Dolly’ shot (M=14.67, SD=5.22) shared an identical mean score, while the ‘Track-out’ shot (M=16.25, SD=6.22) and the ‘Pan’ shot (M=16.67, SD=7.67) had slightly higher scores (Figure \ref{fig8}-left). The ‘Still’ shots exhibited the lowest VRSQ score (M=13.25, SD=3.84). However, according to the one-way ANOVA test, these variations in scores between shots were not statistically significant (F(4,55)=0.67, p\textgreater0.05). The overall range of VRSQ scores across the five shots, from 13.25 to 16.6 (M=15.06, SD=1.64), suggested that the movement designs of the shots were adequately tolerable and easy to follow.

\textit{Immersive Experience:} Conversely, the FilmIEQ results demonstrated statistical significance. The one-way ANOVA test (F(4,55)=7.18, p\textless0.05) with multiple comparisons revealed significant disparities between shots (Figure \ref{fig8}-right). The ‘Track-in’ shot (M=126.33, SD=12.22) achieved the highest mean score, with the 
‘Dolly’ (M=122.5, SD=21.29) and ‘Track-out’ (M=122.0, SD=19.44) shots ranked second and third, respectively. The ‘Still’ shot (M=94.17, SD=14.19) obtained the lowest mean score, while the ‘Pan’ shots displayed a relatively lower performance (M=103.5, SD=21.72). Multiple comparison results showed that the ‘Track-in’, ‘Track-out’ and ‘Dolly’ shots scored significantly higher than the ‘Pan’ and ‘Still’ shots. However, there were no significant differences within the group consisting of ‘Track-in’, ‘Track-out’, and ‘Dolly’ shots, and similarly, between the ‘Pan’ and ‘Still’ shots.

\section{Discussion}
\subsection{Analysis of Integrating Moving Shots in Virtual Reality Sports Broadcasting}
The audio-visual language is a fundamental component of moving shots in the traditional video and allows directors to dictate their creative vision to the production team. This critical aspect translates seamlessly into broadcasting, where the director communicates with the camera crew to capture, transition, and edit various shots. Consequently, we proposed integrating audio-visual language into VR moving shots and explored its benefits relative to the existing still shots, explicitly concerning the audience's viewing experience.

According to our study, several notable findings emerged. Primarily, the duration of exposure to the VR moving camera significantly influences the user's immersion. In Study 1, a single clip scenario, no discernible difference between shots was reported in the FilmIEQ results. However, Study 2 revealed that immersive experiences levels in the case of three moving shots were significantly higher than that of the ‘Still’ shot and remaining moving shots. We posit that this difference primarily arises from the duration of user exposure to the VR environment. Participants in the interviews often expressed that single clips needed longer to elicit a response, despite having been viewed once. Conversely, participants in Study 2 reported that experiencing the narrative of the entire round led to an enhanced sense of immersion. Furthermore, even though participants were exposed to the VR environment for longer durations, the specifically designed VR shots did not significantly contribute to an increased sense of VR sickness.

Secondly, our findings suggest that, within the VR environment, moving shots are generally better received and provide a superior experience compared to still shots. As evidenced in Study 2, significant differences in immersion were noted; the moving camera significantly enhanced the user's viewing experience without inducing VR sickness. This implies that incorporating more moving footage in future VR sports broadcasts could be advantageous. 

Lastly, we found that user preferences and attitudes toward VR broadcasts from different shots are highly individualized. In our brief interviews, participants were asked about their preferred footage type (i.e., the one that provided them with the best experience), but their responses needed a discernible pattern. For instance, one participant preferred the Track-in shot due to their preference for third-person games, while another favored the Pan shot to gain a broader perspective. 
Another interesting finding in the results of Experiment 2 is that the Pan shot significantly differs from the other three moving shots in terms of immersion. Therefore, it is imperative to develop generalized guidance for applying audio-visual language in VR sports broadcasting.

In conclusion, our findings lay the groundwork for an initial theory on VR moving shots in sports broadcasting, thereby opening up avenues for future research. However, we acknowledge that the issue of sample size might pose a limitation to our analysis. Thus, subsequent studies could delve deeper into related VR broadcast applications, the audio-visual language system, and large sample viewer experiences.

\subsection{Design Implications for VR Moving Shots in Broadcasting}
Through the insights gathered from our study, along with our production experiences, we also found some of the design implications for VR moving shots. The production of VR moving shots involves three core elements: orientation, the primary content subject in the first frame, and the VR camera's trajectory. These elements align with those found in conventional film and television production. However, while viewing moving VR shots, viewers' orientation and subject content cannot be confined beyond the initial frame. Unlike 2D media, users in the VR environment possess greater autonomy, which introduces increased uncertainty into the VR narrative. The viewers' potential fear of missing out (FOMO) may result in attention distraction and a reduced sense of presence \cite{Aitamurto2021FOMO}. Consequently, the first frame of every shot in VR broadcasting becomes critical as it is the only initial control point for producers to guide the viewers' experience. Our Study 1 findings illustrate that viewers continue following the main subject presented in the initial frame.

In addition, the moving trajectory of VR shots holds significant narrative value in conveying the story of the entire sports event. Narrative methods are indispensable in time-based media like movies, broadcasts, novels, and games, guiding the viewer through the scene \cite{Godde2018Narration}. The narration of the entire sports event unfolds through sequential shots, each corresponding to a different narrative element. For moving shots in VR, proper guidance and editing methods are required to help the director effectively present the narrative content. This guidance can be incredibly potent during event climaxes, where well-designed moving shots can provide viewers with a significantly immersive VR sports event experience.
Nonetheless, most existing methods and strategies are designed for still VR shots \cite{Moghadam2017transition, Jessica2016Blink, Tina2017Cut}. Drawing upon our experiences and knowledge, we propose that future research could amalgamate the characteristics and theories of existing 2D video and VR still shots. For longer-duration experiences like VR sports broadcasts, the narrative is essential in retaining viewers' attention and interest. Designing moving shots based on different event content could introduce new narrative possibilities in VR media. For instance, directors can employ pan shots to guide the viewer through calmer events or track-in shots to intensify tension during conflicting events.

\subsection{Enhancements and Prospective Developments for VR Sports Broadcasting}
The field of VR sports broadcasting has been evolving over several years. Many professional games (i.e., NHL and NBA) have already adopted VR broadcasting as an optional viewing method. However, our field study and consultations with domain experts have identified several issues that require resolution.

Firstly, current VR broadcasts predominantly rely on still 360-degree shots, lacking the utilization of dynamic 360-degree shots. Our research findings highlight the significant advantages of incorporating moving VR shots, as they enhance viewers' immersion and facilitate the capture of crucial information. Nonetheless, several challenges hinder the production of moving VR shots, including the potential induction of VR sickness during viewing and technical limitations of the equipment, such as the absence of zoom and rotation functionalities. Although conventional equipment like sliders or drones are commonly used in broadcasting, adapting them to different sports stadiums poses considerable implementation challenges and limitations. Volumetric imaging technology, which enables real-time footage capture and 3D space reconstruction, holds promise as a viable solution for VR broadcast technology. However, existing VR moving footage remains insufficient for prolonged, dizziness-free viewing, currently serving primarily as a highlight reel to enrich the viewer experience\cite{Oyman2019Volumetric}.

Additionally, the interactive opportunities provided to audiences within VR sports broadcasts still need to be improved. Our survey reveals that in current VR broadcasts, broadcasters deploy panoramic cameras at specific locations around the stadium, such as the stands, bench, and VIP box. The only interactive option for viewers is switching between different camera perspectives. Thus, We propose that innovative interactive designs could enrich viewers' experiences by providing greater autonomy of choice. For instance, designing an interactive interface within VR for selecting different shots via natural interaction methods, introducing multi-sensory (tactile and olfactory) viewing experiences or adding bullet comments as a communication methods when watching with friends could substantially enhance audience engagement. Nevertheless, we advocate preserving a general viewing method that remains accessible and enjoyable for viewers who may not be well-versed in sports.

Lastly, the current state of VR sports broadcasting lacks a comprehensive, end-to-end production solution. Unlike movies and videos, sports broadcasting combines competitive, entertaining, and unpredictable features \cite{boyle2002new, Turner2007Impact}. Traditional sports broadcasting is divided into pre-game, in-game, and post-game phases, encompassing pre-game previews and commentary, in-game commercials and live feeds, highlight replays, and post-game interviews and commentary. Due to the unique attributes of VR, the original broadcast system and content cannot be directly transferred to VR live broadcasts. Current challenges include devising a specialized system for VR sports broadcasts, tailoring broadcast shots to suit VR characteristics, and enhancing the existing platform's diversity, excitement, and uniqueness of VR event broadcasts. Nonetheless, Wang et al. introduced a video creation tool, ‘Write-A-Video’, that facilitates straightforward video generation by leveraging existing video libraries and simple text input \cite{Wang2019Write}. Consequently, we believe that such technology could soon find application in producing VR sports broadcasts and recaps.

\section{Conclusion}
This paper presented two user studies investigating the impact of moving shots in VR broadcasting, utilizing ice hockey as a representative example. Through our field research, we pinpointed several challenges in the VR broadcasting of ice hockey. To make it feasible, we introduced the concept of event segmentation and developed an ice hockey digital twin environment for the following research.
In Study 1, we asked participants to view single CVR clips featuring five distinct shots (four moving and one still) based on the principles of audio-visual language. Participants then completed questionnaires concerning their sense of immersive experience and any VR-induced discomfort. Our results indicated no substantial difference in viewing experiences between still and moving shots for individual clips.
Proceeding to Study 2, we curated four extended moving shots and one still shot for a given event. We replicated the same evaluation process as in Study 1. Intriguingly, we observed that several moving shots significantly outperformed the still shots regarding viewer experience. 
To sum up, we propose that moving shots offer distinct advantages over still shots in VR broadcasting, providing viewers with a more immersive and comprehensive experience. In light of our investigation and analysis of moving shots in VR ice hockey broadcasts, we have also discussed potential design considerations for VR production and provided suggestions for improving future VR broadcast applications.
\acknowledgments{
We extend our heartfelt appreciation to Xinyi Wang, Yuqi Wu, and Kaige Zhang for their diligent help to this project, as well as to all participants for their time and efforts. This research was supported by Tsinghua University Initiative Scientific Research Program (20213080010), the Foundation of the Ministry of Education of China (22YJCZH041) and the Sichuan Animation Research Center Program of China (DM202213)}

\bibliographystyle{abbrv-doi}

\bibliography{Mybib}
\end{document}